\documentclass[aps,prl,twocolumn,groupedaddress]{revtex4}

\usepackage{graphicx}

\begin{document}

\newcommand{\trS}{\mbox{$^3 \! S _1$}}
\newcommand{\trP}[1]{\mbox{$^3 \! P _{\, #1}$}}


\title{A Large Atom Number Metastable Helium Bose-Einstein Condensate}



\author{A.\ S.\ Tychkov}
\author{T.\ Jeltes}
\author{J.\ M. McNamara}
\author{P.\ J.\ J.\ Tol}
\author{N.\ Herschbach}
\author{W.\ Hogervorst}
\author{W.\ Vassen}
\email{w.vassen@few.vu.nl}
\affiliation{Laser Centre Vrije Universiteit, De Boelelaan 1081, 1081 HV Amsterdam, The Netherlands}


\date{\today}

\begin{abstract}

\noindent We have produced a Bose-Einstein condensate of metastable helium ($^4$He$^\ast$) containing over $1.5 \times 10^7$ atoms, which is a factor of 25 higher than previously achieved. The improved starting conditions for evaporative cooling are obtained by applying one-dimensional Doppler cooling inside a magnetic trap. The same technique is successfully used to cool the spin-polarized fermionic isotope ($^3$He$^\ast$), for which thermalizing collisions are highly suppressed. Our detection techniques include absorption imaging, time-of-flight measurements on a microchannel plate detector and ion counting to monitor the formation and decay of the condensate.
\end{abstract}

\pacs{32.80.Pj, 05.30.Fk, 05.30.Jp, 34.50.Fa}

\maketitle

\noindent Ten years after the first experimental realization of Bose-Einstein condensation (BEC)
in dilute, weakly interacting atomic systems \cite{Anderson95Davis95} the field of degenerate quantum 
gases has developed into a major area of research. For most elements it has not yet been possible to produce Bose-Einstein condensates
containing large numbers of atoms. Only for hydrogen, sodium and rubidium have condensates with more than $10^6$ atoms been realized \cite{Streed05}. Large condensates provide a better signal-to-noise ratio, allow a study of both the collisionless and the hydrodynamic regime and are especially useful for sympathetic cooling and atom laser applications. 
In this realm metastable atoms are of particular interest, offering alternative detection methods due to their high internal energy. Absorption imaging is the technique most frequently applied to detect and measure a BEC. Metastable helium ($2\text{ }^3$S$_1$ state) is the only species for which detection of the condensate has been performed using a microchannel plate (MCP) detector \cite{Robert01}; the same detector was also used to measure ions produced by Penning ionization (while absorption imaging was not used). 
In a recent experiment, after pioneering work with metastable neon \cite{Yasuda}, Schellekens \emph{et al.} used a position sensitive MCP detector to observe the Hanbury Brown and Twiss effect both above and below the BEC threshold \cite{Schellekens}. In the second experiment 
in which BEC was realized with He* \cite{Pereira01}, optical detection 
was used and up to $6 \times 10^5$ atoms could be condensed. Recently, this group reported a high precision measurement of the metastable helium scattering length $a$, which was performed using a two-photon dark resonance \cite{Moal05}. The value $a=7.512\pm0.005$~nm is favorable for experiments with the fermionic isotope $^3$He$^\ast$. It ensures a stable ultracold $^4$He$^\ast$/$^3$He$^\ast$ boson-fermion mixture, as the inter-isotope scattering length will be large and positive \cite{Stas04}. 
Large numbers of $^4$He$^\ast$ atoms all along to the critical temperature provide an efficient 
reservoir for sympathetic cooling and will facilitate the production of degenerate $^3$He$^\ast$ clouds with large numbers of atoms.

In this letter we present an experiment that combines the
various detection methods used previously \cite{Robert01, Pereira01} and describe 
the realization of a BEC of $^4$He$^\ast$ containing more than $1.5 \times 10^7$
atoms. This large improvement is primarily due to the application of one-dimensional Doppler 
cooling inside the magnetic trap rather than three-dimensional Doppler cooling prior to magnetic trapping. Doppler cooling of polarized atoms was originally proposed for atomic hydrogen \cite{Walraven89}, and has recently been demonstrated for optically dense samples of 
magnetically trapped chromium atoms \cite{Schmidt03}. Compared to the laser cooling methods we investigated previously 
\cite{Herschbach03Tol99, Tychkov04}, this configuration is more efficient and simple. 

The experimental setup is an extended and improved version of our previous setup \cite{Herschbach03Tol99}. In short, we start with a beam of metastable atoms produced by a liquid-nitrogen cooled dc-discharge source. The atomic beam is collimated, deflected and slowed by applying laser beams resonant with the $2\text{ }^3$S$_1 \rightarrow 2\text{ }^3$P$_2$ transition at 1083~nm. Typically $2\times10^9$ atoms are loaded into a magneto-optical trap (MOT) at a temperature of 1~mK. Since our previous experiments \cite{Herschbach03Tol99}
we have installed a new ultrahigh vacuum (UHV) chamber and magnetic trap. The coils for our cloverleaf magnetic trap are placed in water-cooled plastic boxes and positioned in re-entrant windows. Inside the UHV chamber two MCP detectors (Hamamatsu F4655) and an rf-coil are mounted. The first MCP detector is positioned $\sim$10~cm from the trap center and attracts positively charged ions produced in Penning ionizing collisions:
$\text{He*} + \text{He*} \rightarrow \text{He}^+ +
\text{He}(\mbox{1 $^1\text{S}$}) + e^-$ (or $\text{He*} + \text{He*}
\rightarrow {\text{He}_2}^{\! +} + e^-$). These ionization processes are the primary loss mechanisms in cold clouds of He*. 
A second (identical) MCP detector shielded by a grounded grid is positioned 17~cm below the trap center and detects neutral He* atoms that fall upon it. This detector is mounted on a translation stage and can be displaced horizontally to allow a vertical laser beam to pass through the trap center for absorption imaging. Absorption imaging of the MOT cloud determines the number of atoms in the MOT with an accuracy of about 20\% \cite{Herschbach03Tol99}; this is used to calibrate the He* MCP detector. 
When the MOT is loaded we switch off all currents and laser beams. In an applied weak magnetic field we spin-polarize the cloud (atoms are pumped into the $m=+1$ magnetic sublevel) and switch on the currents of the cloverleaf magnetic trap. Typically $\sim$60\% of the atoms is transfered from the MOT into the magnetic trap in this procedure. We operate the cloverleaf trap at a bias magnetic field $B_0$=24~G to suppress excitation of depolarizing transitions in the subsequent one-dimensional Doppler cooling stage.  

\begin{figure}
\includegraphics[width=0.9\columnwidth]{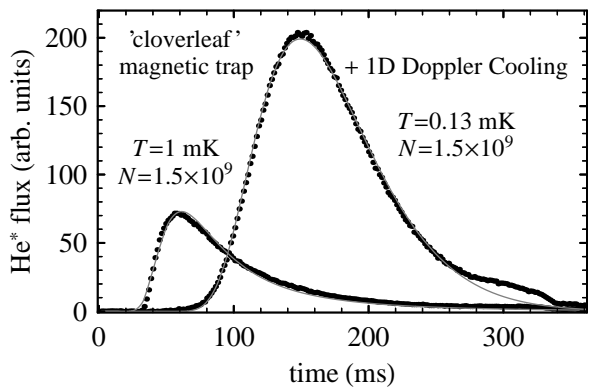}
\caption{ Time-of-flight signals of $^4$He$^\ast$ atoms released from the magnetic trap, with and without
one-dimensional Doppler cooling. The apparent signal increase after Doppler cooling is due to the increased 
fraction of atoms that is detected at lower temperature. The line is a fit assuming a
Maxwell-Boltzmann velocity distribution.
\label{fig1DDoppler}}
\end{figure}

One-dimensional Doppler cooling starts at the same time as the cloverleaf trap is switched on. 
It is implemented by retroreflecting a weak circularly polarized laser beam along the (horizontal) symmetry axis of the magnetic field. 
During the cooling pulse the temperature decreases reducing the size and increasing the optical thickness. Cooling in the radial directions relies on reabsorption of spontaneously emitted red-detuned photons by the optically thick cloud. Other possible energy redistribution mechanisms are collisional thermalization and anharmonic mixing. While the collision rate increases 
from 1.5~s$^{-1}$ to 20~s$^{-1}$ during Doppler cooling, anharmonic mixing is negligible in our trap. With a laser detuning of one natural linewidth below the resonance frequency for an atom at rest in the center of the trap and an intensity of 
$10^{-3} I_{\text{sat}}$ ($I_{\text{sat}}=0.17$ mW/cm$^2$), optimum cooling is realized in 2 seconds. In a separate experiment $10^8$ $^3$He$^\ast$ atoms were loaded into the magnetic trap and cooled using the same technique.
For identical fermions s-wave collisions are forbidden, while the contribution of the higher-order partial waves is highly suppressed in this temperature range for He$^\ast$. We observed a temperature decrease from 1~mK to 0.15~mK, which suggests that reabsorption of red-detuned photons scattered by atoms in the cloud is the main cooling mechanism in the radial direction.

Figure~\ref{fig1DDoppler} shows two TOF traces, illustrating the effect of one-dimensional 
Doppler cooling in our cloverleaf magnetic trap. We typically trap $N=1.5\times10^9$ atoms, which are cooled to a temperature $T=0.13$~mK, three times the Doppler limit. This implies a temperature reduction by a factor of 8 and an increase in phase-space density by a factor of $\sim$600, while practically no atoms are lost from the trap. For comparison, reaching this temperature by means of rf-induced evaporative cooling would result in the loss of $\sim$90\% of the atoms from the trap.
In previous experiments \cite{Herschbach03Tol99, Tychkov04} we applied three-dimensional
Doppler cooling or a two-color magneto-optical trap to improve the starting conditions for evaporative cooling. One-dimensional Doppler cooling provides lower temperatures, higher phase-space density and is easier to implement.  
At this point the lifetime of the atoms in the magnetic trap is about 3 minutes, limited by collisions with background gas. To compress the cloud we reduce the bias field in the trap center to ~3 G in 2.5 seconds, which increases the temperature  to 0.2~mK. The parameters of our magnetic trap then are modest: the axial and radial trap frequencies are $\omega_z/2\pi=47\pm 1$~Hz and $\omega_{\perp}/2\pi=237\pm 4$~Hz respectively. Higher frequencies are possible but are not required to achieve BEC. Our procedure is similar to previous
experiments on evaporative cooling in our group \cite{Tol04}. After compression we cool the gas by rf-induced evaporative cooling in 12~s to BEC, which is achieved at a temperature of $\sim2\mu$K. We apply a single exponential rf ramp, starting at 50~MHz. The frequency decreases to zero but the ramp is terminated at around 8.4 MHz. Shorter ramps, down to 2~s, also produce a condensate, albeit with fewer atoms.

\begin{figure}
\includegraphics[width=0.8\columnwidth]{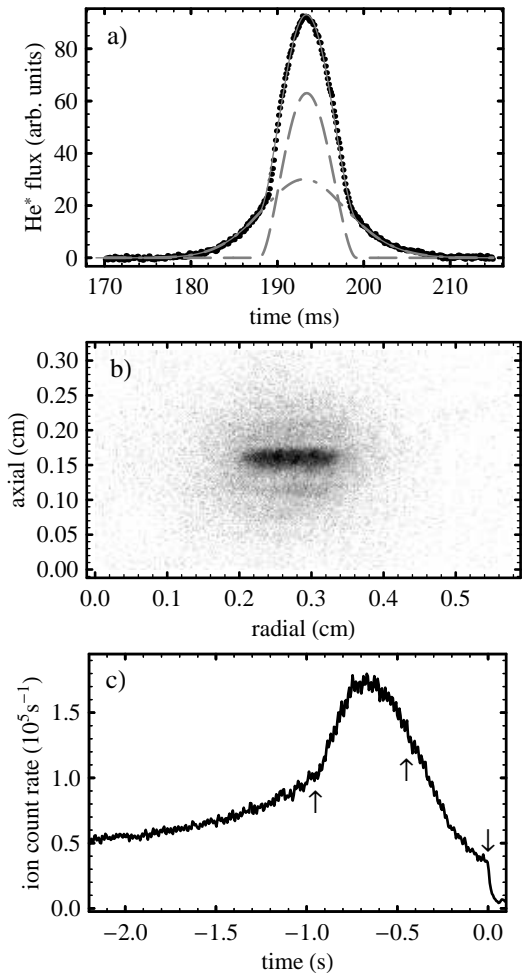}
\caption{Observation of BEC, (a) on the He$^\ast$ MCP detector;
the dashed fit shows the condensed fraction and the dashed-dotted fit the broader thermal distribution,
(b) on a CCD camera; after an expansion time of 19~ms a round thermal cloud surrounding a cigar-shaped condensate is visible, (c) on the ion MCP detector; the condensate starts to grow at $t=-0.95$ s, at $t=-0.45$ rf ramp ends and at $t=0$ the trap is switched off.
\label{figBEC}}
\end{figure}

The most sensitive method to detect BEC is TOF analysis of the expanding cloud on the He$^\ast$ MCP detector.  
A typical TOF signal, obtained in a single shot, is shown in Fig.~\ref{figBEC}a. The double structure is a combination of the inverted parabola indicating the presence of a BEC released from the harmonic trap and a broad thermal distribution. This signal is used to determine the number of atoms in the condensate as well as in the thermal cloud.
In contrast to the Orsay experiments \cite{Robert01, Schellekens}, all atoms stay in the $m=+1$ sublevel during the trap switch-off. Applying the MCP calibration, the area under the fitted curve determines the number of atoms that have hit the detector. When we consider a thermal cloud, this number is only a small fraction of the total number of thermal atoms $N_{th}$. Therefore the determination of $N_{th}$ relies upon the measured temperature and the MCP calibration. The condensate expansion, determined by the mean-field interaction energy, is much slower. Thus the condensate will fall completely on the detector sensitive area (diameter of 1.45~cm), allowing us to measure the number of condensed atoms $N_0$ using the MCP calibration alone. The maximum number of atoms in the condensate deduced in this way is $1 \times 10^7$. This number is an underestimate due to MCP saturation effects, which will be discussed below.
By applying an additional magnetic field pulse we compensate small residual field gradients, which otherwise lead to a slight deviation from free fall expansion. With stronger field pulses we can also push the cloud towards the detector and this way realize shorter expansion times. The model used to fit the time-of-flight signals of the partly condensed clouds (Fig.~\ref{figBEC}a) is a combination of a thermal part (Bose distribution) and a condensate part (parabolic distribution). The chemical potential $\mu$, the number of atoms and the temperature $T$ are the free parameters of the fit; effects of interactions are not included in the function used for the thermal part. In the Thomas-Fermi limit \cite{Dalfovo}, the chemical potential is given by $\mu^{5/2}=15\hbar{^2}m^{1/2}2^{-5/2}N_{0}\bar{\omega}^3 a$, where $\hbar$ is Planck's constant divided by $2 \pi$, $\bar{\omega}$ is the geometric mean of the trap frequencies, $m$ is the mass of the helium atom and $a=7.512(5)$~nm \cite{Moal05} is the scattering length. A maximum value of $\mu$ extracted from the fit of the TOF signal is $\sim1.3\times 10^{-29}$~J, which corresponds to $5.1\times10^7$ atoms in the condensate. A possible cause for the discrepancy between the number of atoms determined from the integrated signal and from the measurement of the chemical potential may be saturation of the MCP when detecting a falling condensate (peak flux of $\sim10^9$ atoms/second); this leads to an underestimation of $N_0$ as well as $\mu$. Another possible cause is distortion of the velocity distribution during the trap switch-off and influence of remaining magnetic field gradients on the expansion of the cloud. This may lead to an overestimation of $\mu$, and therefore also of $N_0$.

When the MCP detector is shifted horizontally, we can detect the condensate
on a CCD camera. A weak ($I=10^{-1} I_{\text{sat}}$), 50~$\mu$s long on-resonance laser pulse is applied to
image the shadow of the atoms on the CCD camera for which a quantum efficiency of $\sim$1.6\% is measured at 1083 nm. As expected, the condensate expands anisotropically while the thermal cloud shows an isotropic expansion (Fig.~\ref{figBEC}b). Absolute calibration of the number of atoms at $\sim\mu$K temperatures could not be performed by optical means. The analysis of the absorption images, taken between 1~ms and 70~ms after the trap was switched off, shows that the condensate expansion deviates from the theoretical predictions \cite{Castin96}: it expands faster than expected in the radial direction and slower in the axial. From these observations we conclude that the expansion of the cloud is influenced by magnetic field gradients during the switch-off of the trap. A difference in the switch-off times of the axial and radial confinement could cause an additional imbalance in the redistribution of the condensate interaction energy between the two directions. This may influence the measurements of both the chemical potential and the temperature. In order to check if the interaction energy is conserved, we extract the asymptotic kinetic energy gained during the expansion from absorption images of the cloud \cite{Holland}. In the Thomas-Fermi approximation this so-called release energy should equal the interaction energy in the trap. We obtain $N_{0}=4\times10^7$ from this analysis assuming that no extra energy is added to (or taken from) the system during the trap switch-off. This is not exactly fulfilled in our case as switching is not fast enough to ensure diabaticity. 

\begin{figure}
\includegraphics[width=0.9\columnwidth]{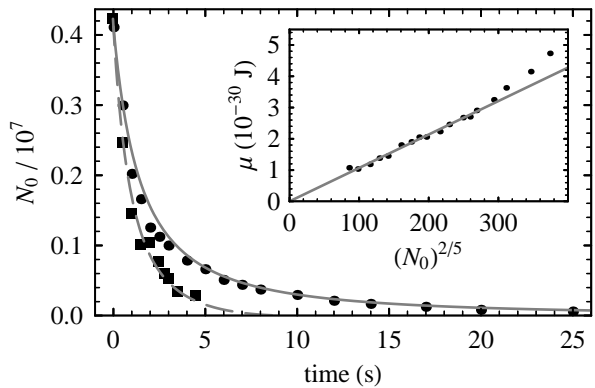}
\caption{Decay of a quasi-pure BEC (circles) and a BEC in the presence of a large ($N_{th}=N_0$) thermal fraction (squares). The dashed curve represents the atomic transfer model \cite{Zin03}, with two- and three-body loss rate constants obtained from a fit (full curve) to the decay of the quasi-pure BEC. Data points that lie above $N_0=10^6$ are corrected for the saturation effects.
Inset: Chemical potential ($\mu$) as a function of $N_0^{2/5}$. The same data as for the quasi-pure BEC decay are used in the plot. The value of $\mu$ is multiplied by a constant (0.61), to bring the data points on the theoretical line (see text).
\label{figDecay}}
\end{figure}

To verify our TOF signal analysis, we plot the chemical potential as a function of $N_{0}^{2/5}$ using data obtained from the MCP measurements (here $N_{0}$ is the number of condensed atoms measured by integrating the MCP current). The data points lie on a straight line which goes through zero with a slope larger than expected, meaning that either $\mu$ is overestimated or $N_0$ is underestimated. The former is supported by the analysis of the absorption images so we correct $\mu$. The corrected data points as well as the theoretical line are presented in the inset of Fig.~\ref{figDecay}. The plot also shows that the MCP detector saturates, when the number of atoms in the condensate exceeds $\sim10^6$. When we now extract the number of atoms from the measured chemical potential, after the correction for the distortion during the trap switch-off, we find $N_0=1.5\times 10^7$ in our largest condensates. This number is still a lower limit, as the analysis assumes that $\mu$ is not affected by saturation of the MCP detector. We, however, measure a reduction in $\mu$ when we push the BEC towards the detector, thus increasing the saturation problem.

With a second MCP detector we observed the growth and decay of our condensate by counting the ions produced during evaporative cooling. Due to the increase in density the ion signal increases, although the number of trapped atoms decreases. When BEC sets in, a sharp increase is expected, indicating the formation of a dense cloud in the centre of the trap \cite{Robert01}. This is demonstrated in Fig.~\ref{figBEC}c, which shows the growth of the condensate as well as its decay. 

The dynamics of formation and decay of the condensate is an interesting aspect that was discussed and investigated earlier to some extent \cite{GrowthDecay}. In our group a model was developed describing the decay of the condensate in the presence of a thermal fraction \cite{Zin03}. The model assumes thermalization to be fast compared to the rate of change in thermodynamic variables, so the system remains in thermal equilibrium during the decay. It was shown that under this assumption a transfer of atoms should occur from the condensate to the thermal cloud, enhancing the condensate decay rate. To verify this, we performed measurements of the BEC lifetime using the TOF signal. Due to the high detection efficiency it was possible to detect a condensate up to 75~s after it was produced. Results of these measurements are summarised in Fig.~\ref{figDecay}. We fit the model to the experimental data for a quasi-pure BEC decay; two- and three-body loss rate constants are used as free parameters. Good agreement with the experiment is found for two- and three-body loss rate constants $\beta=2\times 10^{-14}$ cm$^3$ s$^{-1}$ and $L=9\times 10^{-27}$ cm$^6$ s$^{-1}$ respectively, which compares well with theory \cite{Fedichev96Leo01}. When we use the extracted values of $\beta$ and $L$ in our model for the decay of the condensate in the presence of a thermal fraction, the dashed curve included in Fig.~\ref{figDecay} is obtained. The agreement with the experiment is good, so we can conclude that the model reproduces the data.

To summarize, we have realized a condensate of $^4$He$^\ast$ containing more than $1.5\times10^7$ atoms 
and studied its growth and decay by measuring the ion production rate in situ, observing its ballistic 
expansion by absorption imaging and by recording the time-of-flight signal on an MCP detector. 
The main ingredient that made this large atom number possible is one-dimensional Doppler cooling 
in the magnetic trap. We demonstrated that this technique can also be applied to cool spin-polarized 
helium fermions, where the Pauli principle forbids s-wave collisions. Combining both isotopes in one setup may allow the observation of Fermi degeneracy in boson-fermion mixtures of metastable atoms.

\begin{acknowledgments}
We thank Jacques Bouma for technical support. This work was
supported by the 'Cold Atoms' program of the Dutch Foundation for Fundamental Research on Matter (FOM) 
and by the European Union (ESF BEC2000+/QUDEDIS program and 'Cold Quantum Gases' network).
\end{acknowledgments}


\begin{thebibliography}{99}
\bibitem{Anderson95Davis95} M.~H.~Anderson \emph{et al.}, Science \textbf{269}, 198 (1995);
K.~B.~Davis \emph{et al.}, Phys.\ Rev.\ Lett.\ \textbf{75}, 3969 (1995).
\bibitem{Streed05} E.~W.~Streed \emph{et al.}, arXiv: cond-mat/0507348 (2005).
\bibitem{Robert01} A.~Robert, \emph{et al.}, Science \textbf{292}, 461 (2001).
\bibitem{Yasuda} M.~Yasuda and F.~Shimizu, Phys.\ Rev.\ Lett.\ \textbf{77}, 3090 (1996).
\bibitem{Schellekens} M.~Schellekens \emph{et al.}, Science \textbf{310}, 648 (2005).
\bibitem{Pereira01} F.~Pereira Dos Santos \emph{et al.}, Phys.\ Rev.\ Lett.\ \textbf{86}, 3459 (2001).
\bibitem{Moal05} S.~Moal \emph{et al.}, arXiv:cond-mat/0509286 (2005).
\bibitem{Stas04} R.~J.~W.~Stas \emph{et al.}, Phys.\ Rev.\ Lett.\ \textbf{93}, 053001 (2004).
\bibitem{Walraven89} T.~Hijmans \emph{et al.}, J.\ Opt.\ Soc.\ Am.\ B \textbf{6}, 2235 (1989).
\bibitem{Schmidt03} P.~O.~Schmidt \emph{et al.}, J.\ Opt.\ Soc.\ Am.\ B \textbf{960}, 203 (2003).
\bibitem{Herschbach03Tol99} N.~Herschbach \emph{et al.}, J.\ Opt.\ B \textbf{5}, 65 (2003);
P.~J.~J.~Tol \emph{et al.}, Phys.\ Rev.\ A \textbf{60}, R761 (1999).
\bibitem{Tychkov04} A.~S.~Tychkov \emph{et al.}, Phys.\ Rev.\ A \textbf{69}, 055401 (2004).
\bibitem{Tol04} P.~J.~J.~Tol, W.~Hogervorst, and W.~Vassen, Phys.\ Rev.\ A \textbf{70}, 013404 (2004).
\bibitem{Dalfovo} F.~Dalfovo \emph{et al.}, Rev.\ Mod.\ Phys. \textbf{71}, 463 (1999);
\bibitem{Castin96} Y.~Castin and R.~Dum, Phys.\ Rev.\ Lett.\ \textbf{77}, 5315 (1996).
\bibitem{Holland} M.~J.~Holland \emph{et al.}, Phys.\ Rev.\ Lett.\ \textbf{78}, 3801 (1997).
\bibitem{Fedichev96Leo01} P.~O.~Fedichev \emph{et al.}, Phys.\ Rev.\ Lett.\ \textbf{77}, 2921 (1996); P.~Leo \emph{et al.}, Phys.\ Rev.\ A \textbf{64}, 042710 (2001).
\bibitem{GrowthDecay} S.~Seidelin \emph{et al.}, J.\ Opt.\ B \textbf{5}, 112 (2003); J.~S\"{o}ding \emph{et al.}, Appl.\ Phys.\ B \textbf{69} 257 (1999); C.~W.~Gardiner \emph{et al.}, Phys.\ Rev.\ Lett.\ \textbf{79}, 1793 (1997); M.~K\"{o}hl \emph{et al.}, Phys.\ Rev.\ Lett.\ \textbf{88}, 080402 (2002).
\bibitem{Zin03} P.~Zin \emph{et al.}, J.\ Phys.\ B \textbf{36}, L149 (2003).
\end{thebibliography}
\end{document}